%% file: main.tex
\documentclass[10pt, conference, compsocconf]{IEEEtran}
\usepackage{flushend}
\usepackage{cite}

\usepackage[pdftex]{graphicx}
\graphicspath{{figures/}}
\DeclareGraphicsExtensions{.pdf,.png}

\usepackage{algorithmicx}
\usepackage{algpseudocode}
\usepackage[tight,footnotesize]{subfigure}

\input{commands}

\begin{document}

\title{A Case Study in Coordination Programming:\\
       Performance Evaluation of S-Net vs Intel's Concurrent Collections}

\author{\IEEEauthorblockN{
    Pavel Zaichenkov\IEEEauthorrefmark{1}\IEEEauthorrefmark{4},
    Bert Gijsbers\IEEEauthorrefmark{2}\IEEEauthorrefmark{3},
    Clemens Grelck\IEEEauthorrefmark{3},
    Olga Tveretina\IEEEauthorrefmark{1},
    Alex Shafarenko\IEEEauthorrefmark{1}}
    \IEEEauthorblockA{\IEEEauthorrefmark{1}
    Compiler Technology and Computer Architecture Group, 
    University of Hertfordshire, United Kingdom\\
    \{p.zaichenkov,o.tveretina,a.shafarenko\}@ctca.eu}
    \IEEEauthorblockA{\IEEEauthorrefmark{2}
    Programming Languages Group, Ghent University, Belgium\\
    bert.gijsbers@ugent.be}
    \IEEEauthorblockA{\IEEEauthorrefmark{3}
    Informatics Institute, University of Amsterdam, The Netherlands\\
    c.grelck@uva.nl}
    \IEEEauthorblockA{\IEEEauthorrefmark{4}
    Moscow Institute of Physics and Technology, Russia}}
\maketitle

\begin{abstract}
\input{abstract}
\end{abstract}


\IEEEpeerreviewmaketitle

\input{body}

\bibliographystyle{IEEEtran}
\bibliography{bib/IEEEabrv,bib/biblio}
\end{document}

%% file: commands.tex
\newcommand{\snet}{\textsc{S-Net}}
\newcommand{\front}{\textsc{Front}}

\newcommand{\langc}{\textsc{C}}

\newcommand{\langsac}{\textsc{SaC}}

\newcommand{\pdffig}[5]{
  \begin{figure}[#2]
  \centering
  \includegraphics[#3]{#1}
  \caption{#4}
  \label{#5}
  \end{figure}
}

%% file: abstract.tex
We present a programming methodology and runtime performance case study 
comparing the declarative data flow coordination language \snet{} with Intel's 
Concurrent Collections (CnC). 
As a coordination language \snet{} achieves a near-complete separation of 
concerns between sequential software components implemented in a separate
algorithmic language and their parallel orchestration in an asynchronous
data flow streaming network.

We investigate the merits of \snet{} and CnC with the help of a relevant 
and non-trivial linear algebra problem: tiled Cholesky decomposition.  
We describe two alternative \snet{} implementations of tiled Cholesky
factorization and compare them with two CnC implementations, one with 
explicit performance tuning and one without, that have previously been 
used to illustrate Intel CnC.
Our experiments on a 48-core machine demonstrate that \snet{} manages to
outperform CnC on this problem.  

%% file: body.tex
\section{Introduction}

The main challenges in the design of concurrent programs are the correct
sequencing of interactions between computational threads, and the control of
shared resources.  Two research directions have been pursued to address these
challenges:
1) new programming models and parallel programming language abstractions;
2) specification of concurrent behavior and automatic generation of concurrent
   code.  In the latter direction, concurrent control code is generated
   automatically based on a specification.  Habermann's path expressions is an
   example of an early work in this direction \cite{AndlerPOPL79,HoareCASM74}.

Yet a third way is to use a {\em coordination language} for the part of the
program that manages concurrent {\em components}.  The components are the
building-block algorithms that the overall algorithm uses in order to produce
the required results.  Each component is concurrency agnostic while the
overall algorithm is, in fact, a concurrent composition of the building blocks.
In the coordination language one specifies the relations between components:
what data are communicated between them and where synchronization takes place.
A macro-dataflow coordination language, such as \snet{}
\cite{GrelSchoShafPPL08,GrelSchoShafIJPP10}, emphasizes communication between
the components as it contains a complete set of wiring primitives that promote
a view of an application as a {\em streaming network} of
\emph{asynchronous components}.

The first such attempt at coordination, albeit in a concurrency-centric, rather
than communication-centric flavor, was by Gelernter and Carriero
\cite{GeleCarrCASM92}, who defined a concurrency/synchronization control
language Linda as a set of primitives to be used with conventional imperative
languages as pseudo-intrinsic functions.  Further attempts brought into focus
the matters of software engineering (in particular abstraction, encapsulation
and inheritance) and the concept of compositionality, i.e. the ability of the
concurrent glue to seamlessly integrate the components into a single program.

The coordination language \snet{} takes the above issues on board.  It uses
streams as a glue with which it connects components into a single application.
The way the components are connected depends on data types, which play a dual
role in \snet{}: they ensure that the correct collection of objects is received
by each of the components in every act of communication, but they also help to
route messages to their destinations through a very small and simple
set of wiring patterns called {\em combinators}.  Structuring the application
into a hierarchical network helps to apply these simple tools systematically.
The initial design is refined by progressively revealing the structure of
subnetworks until the design process stops at individual components.

This paper reports on a new performance study of the language.  As an example
application we choose tiled Cholesky decomposition, a linear algebra algorithm
that lends itself easily to parallelization for a multi-core system.  It
decomposes a Hermitian, positive-definite matrix into the product of a lower
triangular matrix and its conjugate transpose.  For such matrices, the
algorithm is roughly twice as efficient as the more general LU decomposition.
Here we use a tiled version of the procedure originally described by Buttari,
Langou, Kurzak and Dongarra\ \cite{ButtLangKurz+PC09}.  The choice was
motivated by the fact that the algorithm is well used in computational linear
algebra as well as having a sufficiently complex and varied internal structure,
which would benefit from component coordination as a method of concurrent
implementation.  Furthermore, tiled Cholesky factorization has repeatedly been
used to illustrate the merits of Intel's Concurrent Collections (CnC), and
various CnC implementations have been made available. This ensures a reasonably
fair performance comparison.

{\bf Contribution.} We compare the performance of \snet{} with that of
Intel's Concurrent Collections (CnC) \cite{BudiChanKnob+fk09}.  The choice of
CnC for comparison is motivated by the fact that CnC follows a similar
dataflow approach (with some control flow facilities as well) as \snet.
CnC is implemented as a C++ library (but further implementations exist), rather than
a fully fledged coordination language.  It is quickly gathering momentum in both
industrial and academic research.  \snet{}, in contrast, emphasizes
modular parallel program design by means of
strict separation of domain-specific programming from concurrency engineering.
Various \snet{} applications have been built by our industrial partners,
including Thales Research, SAP and Philips Healthcare
\cite{PenHerGre+ICCS10,GrelSchoShafIPDPS07}.

Our second contribution is a declarative specification of a dataflow algorithm
for tiled Cholesky decomposition in \snet{}. This specification avoids two acts of
barrier synchronization, which occur in the imperative version of the
algorithm.  Our measurements show a superior performance of this declarative
version over other three implementations we investigated.

\section{Related Work}

A coordination language describes communication between independent
single-threaded computational processes.  It is responsible for managing
asynchronous components as well as for supporting synchronization and
communication among them \cite{GeleCarrCASM92}.  The coordination language is
unaware of activities inside components, and thus can be added to almost any
language.

A separation of concerns is an important property of coordination languages.
If coordination is separated from computation, these activities can be
implemented in different languages.  The separation of concerns facilitates
different roles for different programmers.  Domain experts can concentrate on
a domain-specific problem while parallel programming experts take care of
concurrency aspects.  The separation of concerns facilitates economy and
flexibility as well.  Components become more generic and thus can be reused in
different contexts.

Linda is the first coordination language \cite{linda}.  Communication between
independent processes is performed using shared memory referred to as tuple
space (elements of the memory are tuples, not bytes).  Elements of tuple space
are accessed by logical name. Therefore, the only information the processes
share is a protocol on element tags.  The separation of concerns in Linda is
not complete because synchronization is located in the computational part of
a program.

Kahn's model of Process Networks (KPN) introduces streams represented as
sequential data channels with infinite capacity to glue independent processes
into a network \cite{kahn74}. The KPN model is a clean coordination model
because coordination does not interfere with the computations inside network
vertices.  \snet{} can be seen a refinement of the KPN model: it takes
engineering aspects, such as memory limitations into consideration. Separation
of concerns in \snet{} is quite clean as components are completely unaware of
the context and synchronization is not interfering with computational activity.

Both \snet{} and CnC use coordination glue to combine separate components
together.  CnC has been significantly influenced by Linda \cite{cncreport},
whereas \snet{} is based on the KPN model.  We are going to show that both
approaches work reasonably well to achieve excellent performance on
a shared-memory parallel platform.

\section{Overview of \snet{}}
\label{sec:snet-overview}

\subsection{The Parallel Component Technology}

Decomposition and encapsulation are general software engineering principles not 
limited to parallel computing.
Problem decomposition results in a representation of an application as a set of
black-box components, whose functionality is defined in terms of the interface
description, with some glue code that holds the components together in a way
that ensures the expected system behavior.

The only requirement to be satisfied by an \snet{} box implementation is the
absence of persistent internal state.  
Stateful components could
neither be moved or cloned in a multicore system, since the new copy would have
to rely on the previous state, which is internal, and hence unavailable.
However, state is somehow required in order to be able to merge messages from
different channels.  Thus, state may only be expressed as a dedicated language
construct at the coordination level. Therefore, state in \snet{} is always
fully explicit.

The consequence is to structure and manage state transitions in the component
world in the same way as control flow is structured and managed in ordinary
programming.  User-defined components become pure functions that map a tuple of
parameters onto a similar collection of results.

As soon as the latter is produced, the internal state should effectively be
destroyed.  Such components are easy to reason about and debug, they are
inherently mobile, and usable as a black box in a parallel computing
environment -- but there is also a price to pay.  The glue environment has to
provide sufficient scaffolding to support an evolving state (or local states!)
of the computation.  In other words, it will need to hold the effective state
of one or more components for them and present it back to the components'
inputs in combination with any data to be processed.

\subsection{The \snet{} Language}

The language \snet{} supports coordination programming by
instantiation of components as {\em boxes} and connecting them by anonymous
data streams \cite{GrelSchoShafIJPP10}. An \snet{} application is represented
as a network between the input and the output, which are two external streams
connecting the whole application with its environment.  In the following 
we briefly revisit the main concepts of the language.

{\bf The box concept.}  A component is instantiated as a Single-Input,
Single-Output (SISO) box.  The box has a limited life cycle: it accepts one
item from the input stream, does some processing and yields zero or more items
to the output stream, after which it destroys its internal state and waits for
the next input item to arrive.  Components are written in a box language, using
the \snet{} communication API.  \langc{} and \langsac{} \cite{GrelSchoIJPP06}
are currently supported as box languages.

{\bf Synchronization.}  In \snet{} the only component which can store and
combine state is the {\em synchrocell}.  For example, the expression
$[|\{r\},\{s\}|]$ synchronizes precisely two messages: one whose type contains
at least a field $r$, another with at least $s$.  All other messages remain
untouched and are forwarded further.  The semantics of the synchrocell is
sequential and does not involve any data transformation, hence concurrency
and mobility concerns do not apply.

{\bf The streaming data concept.}  All boxes accept records as units of their
input. A record in \snet{} is a set of {\em fields} and {\em tags}.  Both
fields and tags have names and values. Field values cannot be examined in
\snet{}: they are references to data which are private to the box language.
Tags are standardized as integers and their values are available in both the
box language and the \snet{} language.
Records are nonrecursive in the sense that it is not possible to define an
unlimited linked structure, such as a list.  

Every user-defined component
contains a program unit (a function or similar) written in a box language, and
a type signature written in \snet{} that defines the type of records (in terms
of their field/tag name sets) that the box accepts and, in a similar way, the
types of any output records that may be produced.
Streams between boxes are sequences of records.  Even though all boxes are
SISO, the data relationships between them are not one-to-one, since streams can
be split and merged using combinators.

\begin{figure}[!t]
  \begin{center}
  \subfigure[]{
    \raisebox{0.9\height}{
    \includegraphics[width=0.2\textwidth]{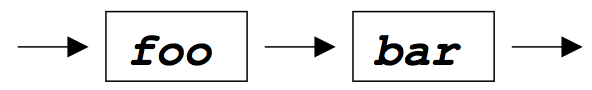}}
    \label{fig:serial}}
  \subfigure[]{
    \includegraphics[width=0.2\textwidth]{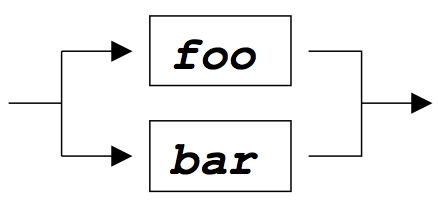}
    \label{fig:parallel}}
  \subfigure[]{
    \raisebox{0.2\height}{
    \includegraphics[width=0.2\textwidth]{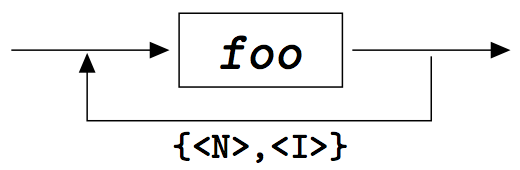}}
    \label{fig:feedback}}
  \subfigure[]{
    \raisebox{0.25\height}{
    \includegraphics[width=0.2\textwidth]{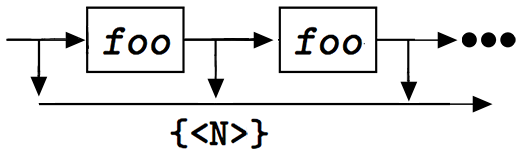}}
    \label{fig:star}}
  \subfigure[]{
    \includegraphics[width=0.2\textwidth]{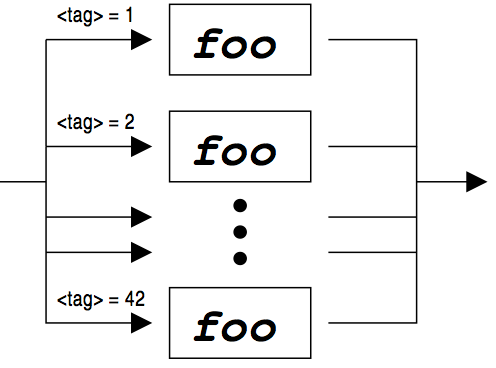}
    \label{fig:split}}
  \caption{Illustration of network combinators used in Cholesky decomposition
    implementation for \snet{}:
    serial combinator (serial composition) \protect\subref{fig:serial},
    parallel combinator (parallel composition) \protect\subref{fig:parallel},
    feedback combinator (feedback loop) \protect\subref{fig:feedback},
    star combinator (dynamic replication) \protect\subref{fig:star} and
    split combinator (parallel replication) \protect\subref{fig:split}.}
  \end{center}
\end{figure}

{\bf Combinators.}  These are second-order functions that ensure
compositionality of SISO networks.  First of all there are serial and parallel
combinators, $A..B$ and $A|B$, respectively.  The serial combinator ``..''
(Fig.~\ref{fig:serial}) connects the output of operand $A$ to the input of
operand $B$, with the input of $A$ and the output of $B$ becoming those of the
resulting network.  The parallel combinator ``$|$'' combines its operands in
parallel, see Fig.~\ref{fig:parallel}.  Incoming records are sent to the
operand network that best matches its type \cite{GrelSchoShafPPL08}.  Type
specifications for complex networks are inferred automatically by the \snet{}
compiler.

The feedback combinator examines the output records of an operand network and
redirects those records that match a pattern back into the input of that
network. For instance, $C{\backslash}z$ creates a feedback loop around operand
network $C$ for records for as long as their type matches type pattern $z$.
This allows a single operand network $C$ to repeatedly process a record until
it is finally converted into something else, see Fig.~\ref{fig:feedback}.

There exist combinators for dynamic replication of a SISO network.  The
expression ${A*x}$ will serially replicate the operand network $A$ an
unspecified number of times (Fig.~\ref{fig:star}).  Only records whose type
matches the given type pattern $x$ escape this network. Thus, the expression is
equivalent to an infinite serial expansion $A..A..A..$ for records as long as
they do {\em not} match exit pattern $x$.  Typically, at some point in time,
due to processing by network $A$, they {\em do} match the exit condition and
will then appear on the output stream of the expression.

Similarly, the expression $B!\textrm{\textless}y\textrm{\textgreater}$
replicates the SISO network $B$ an unspecified number of times in parallel
(Fig.~\ref{fig:split}).  For every unique tag value $y$ one parallel branch
of $B$ is created.  This can be seen as an infinite parallel expansion
$B_{y_0}|B_{y_1}|B_{y_2}|\dots$ The branches persist.  Namely, records with the
same tag value $y$ take the same branch.  All records which encounter this
expression are required to have a type which contains tag
$\textrm{\textless}y\textrm{\textgreater}$.  This is checked at compile time.

Stateful computations can be modelled in \snet{} with an expression like
$([|\{r\},\{s\}|]*\{r,s\} .. MyBox){\backslash}\{s\}$, where a single state
$\{s\}$ is first combined with an incoming record $\{r\}$. Processing by
component $MyBox$ may generate any number and type of output messages, but at
least one evolved state $\{s\}$. The feedback combinator ``$\backslash$'' only
redirects the state back into the network where the process repeats itself with
the next incoming record $\{r\}$.
For details regarding stateful streaming networks in \snet{} we refer the
interested reader to \cite{GrelckIPDPS11}.

\section{Case Study: Cholesky Decomposition}
\label{sec:case-study}

\subsection{The Algorithm}

Cholesky factorization computes a solution to the following problem:  given
a symmetric positive definite matrix $A$, find a lower-triangular matrix $L$,
such that $A = LL^T$.  We use the tiled version of the Cholesky decomposition
algorithm described by Buttari et al.\ \cite{ButtLangKurz+PC09}.

Initially, the input matrix $A$ is decomposed into $p \cdot p$ blocks $A_{ij}$
of size $b \times b$ each. Then, we solve the Cholesky decomposition problem
for all blocks from submatrix $A_{i0}$ separately.  Next, we recompute all
element values in the submatrix $A_{ij}$ (where $i \in (1, p-1), j \in (1, i)$)
and run the algorithm recursively on the submatrix.

\begin{figure}
\begin{algorithmic}[1]
    \For{$k = 0,\ldots,p-1$}
        \State InitialFactorization($A_{kk}, L_{kk}$)
        \ForAll{$j \in (k+1,\ldots,p-1)$}
            \State TriangularSolve($L_{kk}, A_{jk}, L_{jk}$)
        \EndFor
        \ForAll{$j \in (k+1,\ldots,p-1)$}
            \ForAll{$i \in (k+1,\ldots,i)$}
                \State SymmetricRankUpdate($L_{jk}, L_{ik}, A_{ij}$)
            \EndFor
        \EndFor
    \EndFor
\end{algorithmic}
\caption{Tiled Cholesky decomposition algorithm.}
\label{alg:cholesky}
\end{figure}

An overall algorithm is given in Fig.~\ref{alg:cholesky}.  The computational
process is divided into three steps:

{\bf Initial Factorization.} A scalar Cholesky decomposition algorithm is used
to solve $A_{kk} = L_{kk}L_{kk}^T$ equation on this stage.  The result of
computation is a lower-triangular matrix tile $L_{kk}$.

{\bf Triangular Solve.} During this phase we apply the result of the previous
step's computation to solve the equation $A_{jk} = L_{jk}L_{kk}^T$.  The result
is a matrix tile $L_{jk}$.  This step can be performed for all tiles in the
same column concurrently.

{\bf Symmetric Rank Update.} This step is used to update values of tiles
$A_{ij}$, where $j$ ranges from $k+1$ to $p-1$ and $i$ from $k+1$ to $j$. This
is done using the following formula: $A_{ij}' = A_{ij} - L_{ik}L_{jk}^T$.
Similar to the previous step, this can be done concurrently for all $i$ and
$j$.

This numerical problem thus boils down to three building blocks.  Our task in
coordination programming is to glue them together into one coordination program.


\subsection{Implementation with Intel Concurrent Collections}

A detailed description of the CnC implementation of the Cholesky decomposition
algorithm is found in \cite{ChanKnobVuduIPDPS10}.  Here we present only a brief
description of the  CnC concepts relevant to our implementation.

The CnC model has the following important feature: it decouples the
specification of a computation from the expression of its parallelism.
Consequently a domain expert determines the design of the algorithm, and
a tuning expert can be called upon to deal with parallelism, communication,
scheduling and distribution issues, not dissimilar from \snet.

\pdffig{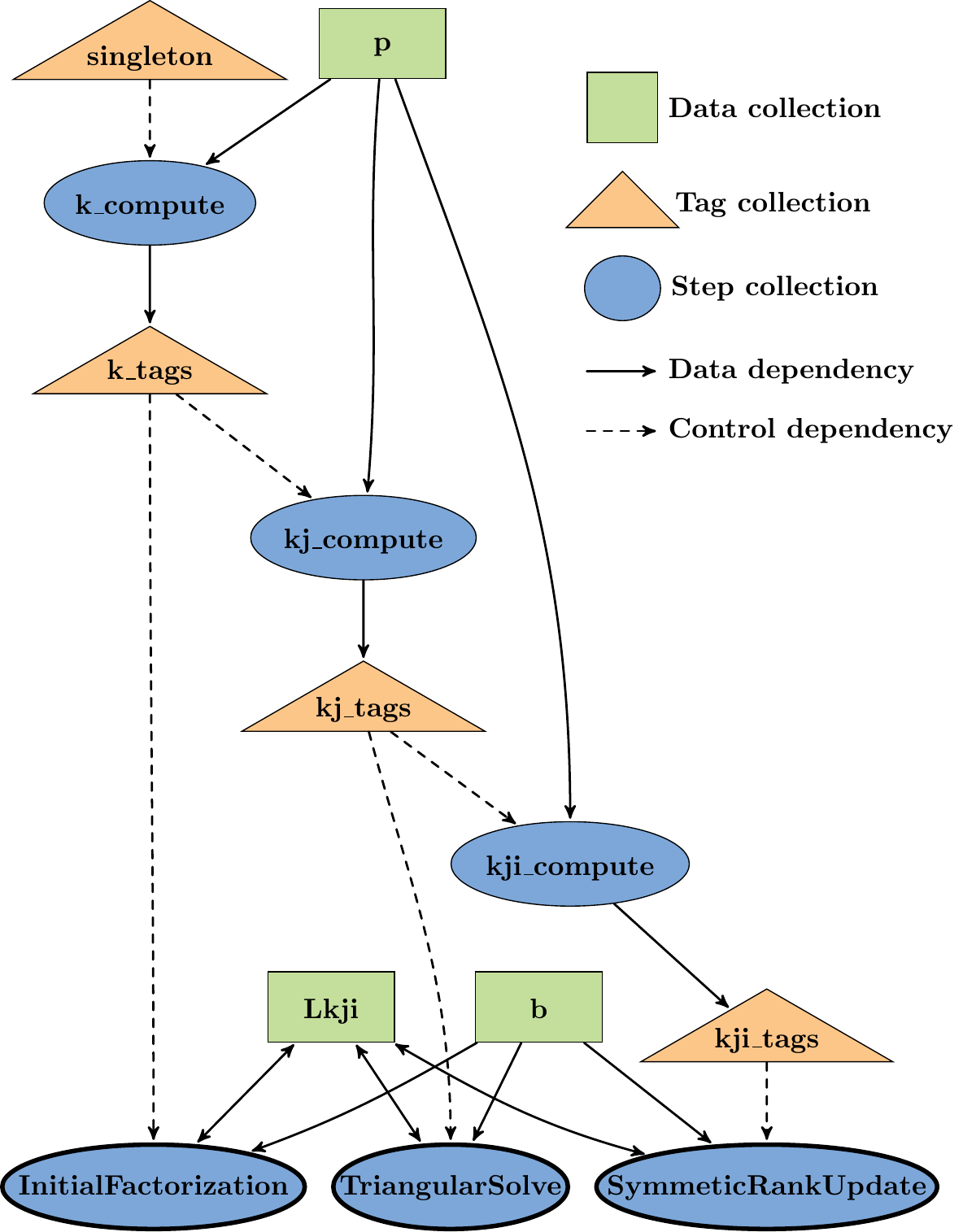}
       {!t}
       {width=3.3in}
       {A computational graph of the CnC implementation of the Cholesky
        decomposition algorithm}
       {fig:cnc_graph}

The domain expert specifies the computation in graph form as depicted in
Fig.~\ref{fig:cnc_graph}.  The graph contains the following types of nodes.

\begin{itemize}

\item A computational {\em step}.  It is a basic unit of execution specified
    explicitly by the domain expert.  In Fig.~\ref{fig:cnc_graph} ellipses
    represent {\em step collections}, which are static declarations of sets
    of dynamic instances.

\item A data {\em item}.  {\em Item collections}\ are used to represent data.
    Items  are elementary units of storage, communication, and synchronization.
    In Fig.~\ref{fig:cnc_graph} there are three item collections shown as
    rectangles: $Lkji$ stores both input matrix $A$ and output matrix $L$, $b$
    stores a block size  and  $p$ is used to calculate the total number of
    blocks in the initial matrix (which equals $p \times p$).

\item A control {\em tag}. Each instance of a step or item has a unique tag,
    which is a tuple of {\em tag components}.  Tags indicate {\em whether}\
    a step will execute, but not {\em when}\ it executes.  A step may produce
    tags as well.  A step collection is associated with exactly one tag
    collection.

\end{itemize}

The relations between steps and items are shown by directed solid edges in
Fig.~\ref{fig:cnc_graph}. They have the following meaning: The line ``item
$\to$ step'' indicates that the step consumes the item and the line ``step
$\to$ item'' means that the step produces the item.

The control relation between a tag collection and a step collection is shown by
directed dashed edges.  A step collection is
associated with exactly one tag collection; multiple step collections may
be {\it prescribed}\ by the same tag collection.

Fig.~\ref{fig:cnc_graph} contains tags and items that do not have inbound
edges.  This means that they are taken from the {\em environment}, which is the
external code that invokes the computation.  For our example the environment
provides $Lkji$, $b$ and $p$ item collections and the \verb$singleton$ tag
collection.

There are three main algorithms given as separate steps in the implementation:
\verb$InitialFactorization$, \verb$TriangularSolve$ and
\verb$SymmetricRankUpdate$ written by the domain expert.  Their behavior is
defined by data collections $Lkji$, $b$ and $p$ they receive as an input data.
Tag collections \verb$k_tags$, \verb$kj_tags$ and \verb$kji_tags$ control the
behavior of each step collection.  All of these steps produce the result of the
computation and put it back into the data collection $Lkji$.

Semantics and execution of CnC are defined as follows.

\begin{itemize}

\item The item or tag is said to be {\em available}\ if it was produced by a
step.

\item The step is said to be {\em prescribed}\ if it was prescribed by a tag
collection and a particular tag became {\em available}.

\item The step becomes {\em inputs-available}\ if all items for this step are
available.

\item The step is {\em enabled}\ and may execute if it is both inputs-available
and prescribed.

\item The program terminates when no step is executing and no unexecuted step
is enabled.  The termination is valid if all prescribed steps have been
executed.

\end{itemize}

Nevertheless, the relation between step, data and tag collections is defined
statically, there is no way to perform scheduling and resource management
dynamically.  In order to solve this shortcoming CnC offers 
a tuning mechanism in the form of special annotations for compiler and 
runtime system.

In order to evaluate the effect of tuning, we evaluate two alternative CnC
implementations of tiled Cholesky factorization.  
The first one does not rely on the tuning, whereas the
second one uses dependency functions in order to improve scheduling.
Dependency functions map control tags into data collection indices in order to
improve scheduling and eliminate stalls during run-time.  Without this
information it is impossible to determine which elements are going to be
accessed by steps, therefore steps are forced to stall.

\subsection{Implementation with \snet{}}

Separation of concerns is also a key feature in \snet{}.  \snet{} describes the
coordination behavior of networks of asynchronous components and their orderly
interconnection via typed streams.  The component implementation is done using
an external {\em box language}\ and \snet{} is not bound to any specific one.

\snet{} uses a message-driven communication model.  The domain expert only
specifies the input and output types of boxes. This is the type signature of
a box. The type signature is basically a declaration of how the type of the
input message is mapped onto the types of the output messages.  On each
computational step a box receives only a single message as input, yet the
number of output messages is not limited.  A box is stateless and runs 
asynchronously with other boxes.  Boxes do not carry global state and
the {\em purity}\ of a function inside a box is a requirement.  A box
computation may start as soon as a message from the input stream is received.

We developed two different implementations for Cholesky decomposition in
\snet{}.  The first one strictly adheres to the algorithm shown in
Fig.~\ref{alg:cholesky}, and thus implements three consecutive steps.
The drawback of this approach is that in each iteration two
barrier synchronizations take place.

Our second implementation is free from barrier synchronization and, hence, 
exposes a higher degree of concurrency. Here, all computations are completely
data-driven as in a dataflow approach: any computational step in the program is
able to execute as soon as the required input data become available.  We now
cover both implementations in more detail.

\subsubsection{Implementation with barrier synchronization}

Compared to CnC, a coordination network in \snet{} does not specify a control
flow.  Box computations depend only on the availability of input data.  Data
relations are completely defined statically by means of type specifications.
In contrast to the CnC graph, the \snet{} network is hierarchically structured.
Any network can structurally play the role of a box in a higher-level network.

\begin{figure}[!t]
  \begin{center}
  \includegraphics[width=3.3in]{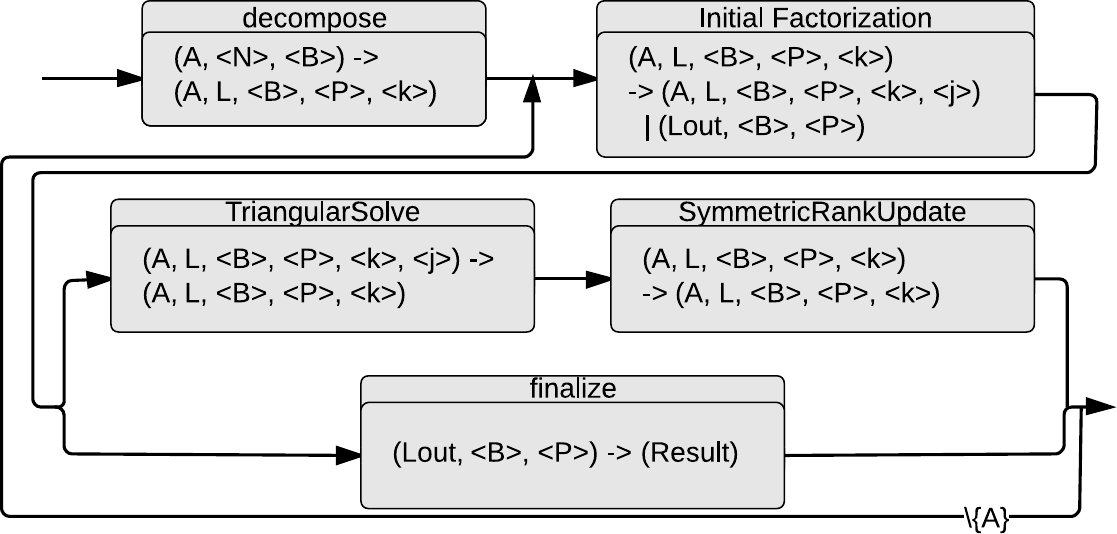}
  \end{center}
  \caption{The \snet{} network for tiled Cholesky decomposition 
   with barrier synchronization.}
  \label{fig:snet_graph}
\end{figure}

In Fig.~\ref{fig:snet_graph} we illustrate the \snet{} network for the first
Cholesky decomposition implementation.  Each box shows its type signature, with
one input type and one or more output types.  {\em Tags} are distinguished from
{\em fields} by angular brackets.  The coordination layer is
able to access tags and route messages based on (integer) tag values. 

The \verb$decompose$ box receives a message with the input matrix $A$, its size
$N$ and the block size $B$ from the environment.  It reallocates the array
where the input matrix is stored and permutes the matrix elements there in
order to improve spatial and time locality.  As the result, box
\verb$decompose$ outputs the input matrix with permuted elements $A$, the
output matrix $L$ filled with zeros (on each iteration of feedback loop
combinator we add new values to the matrix), the block size $B$, the number of
blocks $P$ and an additional iteration index $k$ with initial value zero.

Next, we perform recursive computations (the outer loop in
Fig.~\ref{alg:cholesky}).  The recursion is expressed using a feedback loop
combinator.  The combinator redirects the output of the
\verb$SymmetricRankUpdate$ box to the input of the \verb$InitialFactorization$
box as long as the message containing a field of type \verb$A$ is produced.
Execution terminates once \verb$SymmetricRankUpdate$ stops producing new
messages.  The result of computation is stored in a message that is produced by
\verb$finalize$ box and is sent to the output stream.

We compare the loop index $k$ with the number $P$ in order to determine whether
all the blocks have been computed.  Depending on the result, messages of
different types are sent.  If $k$ is still less than $P$,  we add computed
elements to the matrix $L$ and supply the input record with the additional
index $j$.  \verb$TriangularSolve$ and \verb$SymmetricRankUpdate$ boxes perform
the computations of the corresponding stages.  Lastly, the \verb$finalize$ box
completes the computation by converting the result matrix to a format suitable
for output and performing memory deallocation.

\subsubsection{Data-driven implementation}

\begin{figure}[!t]
  \begin{center}
  \includegraphics[width=2.8in]{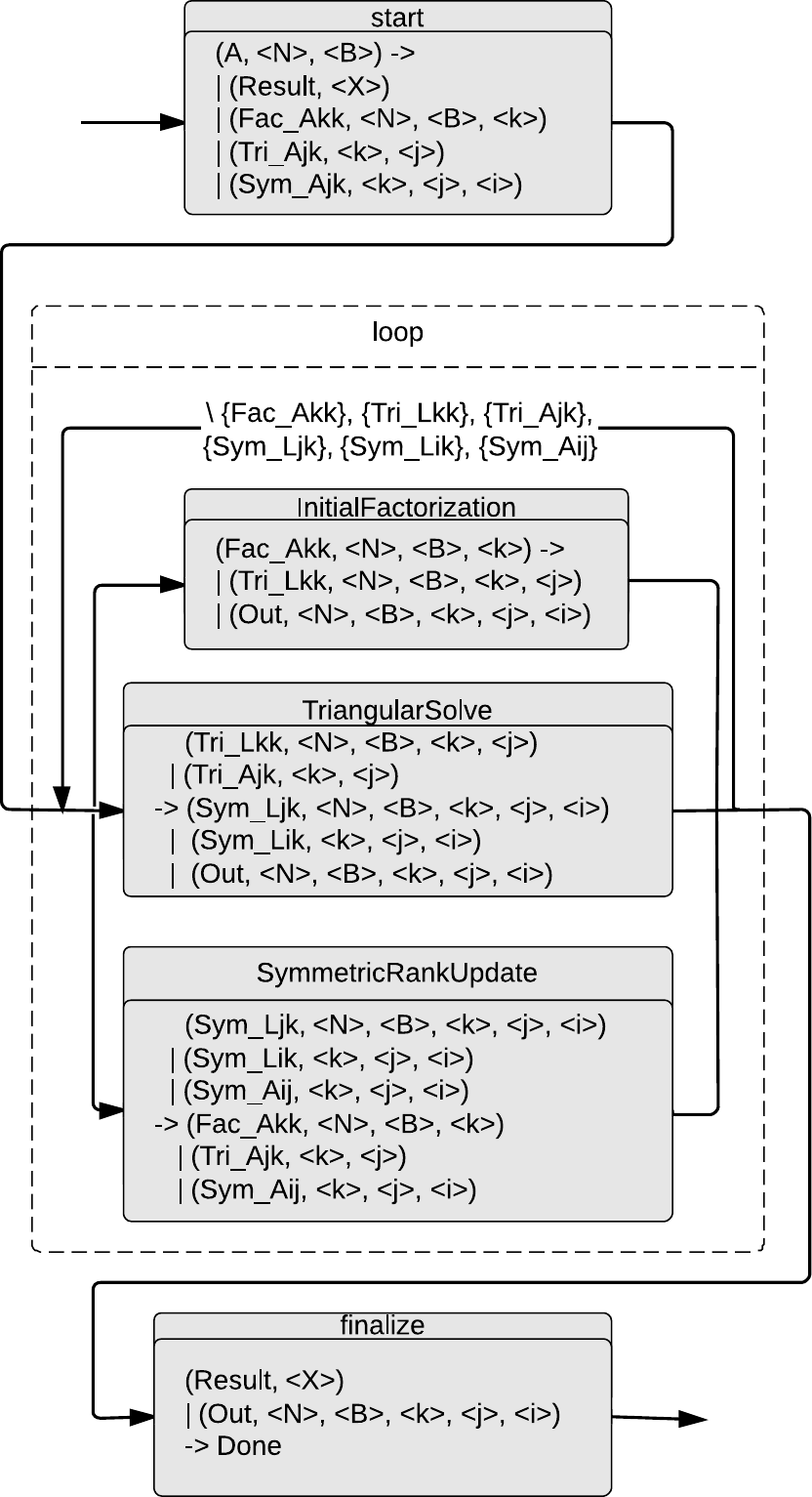}
  \end{center}
  \caption{The \snet{} network for tiled Cholesky decomposition 
   following a purely data-driven approach.}
  \label{fig:snet_graph2}
\end{figure}

In our second \snet{} implementation of Cholesky decomposition the three
central steps from the algorithm are all run in parallel, see
Fig.~\ref{fig:snet_graph2}.

Box \verb$start$ receives the same input, consisting of matrix $A$, matrix size
$N$ and the block size $B$.  It produces one record containing the matrix
$Result$ which is filled with zeros, together with a tag $X$ which denotes the
number of $Out$ tiles which need to be merged with the result matrix in order
to construct the final output matrix.  This $Result$ message is immediately
forwarded to the \verb$finalize$ box. In addition, the \verb$start$ box
produces initial messages containing separate tiles each of which corresponds
to some stage in the algorithm.  That is, it produces messages with diagonal
tiles with type \verb$Fac_Akk$ which are accepted by
\verb$InitialFactorization$; messages with type \verb$Tri_Ajk$ and tag $j$
which are accepted by \verb$TriangularSolve$; and messages with type
\verb$Sym_Aij$ and tags $j$ and $i$ which are accepted by
\verb$SymmetricRankUpdate$.  Tag $j$ denotes a column index and tag $i$ a row
index.  In addition each message contains a loop iteration tag $k$ in order to
distinguish different tiles from different iterations.  All individual box
components are replicated in parallel by the ``$!$'' combinator with index
values taken from the $i$, $j$ and $k$ tags. This ensures that no messages are
queued up in streams waiting for preceding messages to be processed. Input
dependencies can therefore quickly be resolved and computations can start as
soon the input data becomes available.

Messages are transferred between the three central components until result
tiles with type $Out$ are produced which are accumulated by the \verb$finalize$
box.  To this end the \verb$finalize$ is paired with a repeated synchrocell and
both are enclosed in a dedicated feedback loop which is not shown in
Fig.~\ref{fig:snet_graph2}.  Components \verb$TriangularSolve$ and
\verb$SymmetricRankUpdate$ also contain a synchrocell inside.  The synchrocell
in the first component awaits for two messages with \verb$Tri_Ajk$ and
\verb$Tri_Lkk$, representing tiles $A_{jk}$ and $L_{kk}$ respectively.
Similarly, synchrocells in the second component awaits for three tiles
$L_{jk}$, $L_{ik}$ and $A_{ij}$ which are packed into \verb$Sym_Ljk$,
\verb$Sym_Lik$ and \verb$Sym_Aij$ respectively.  Once the dependencies are
satisfied, a box computation starts, which results in one or more output
messages.

From this description it becomes clear that synchronization is only needed to
satisfy specific input dependencies in preparation for individual activations
of box computations.  Collective thread synchronizations are
completely avoided.  Messages are transferred independently and computations
are free to start as soon as all input dependencies are satisfied.

\pdffig{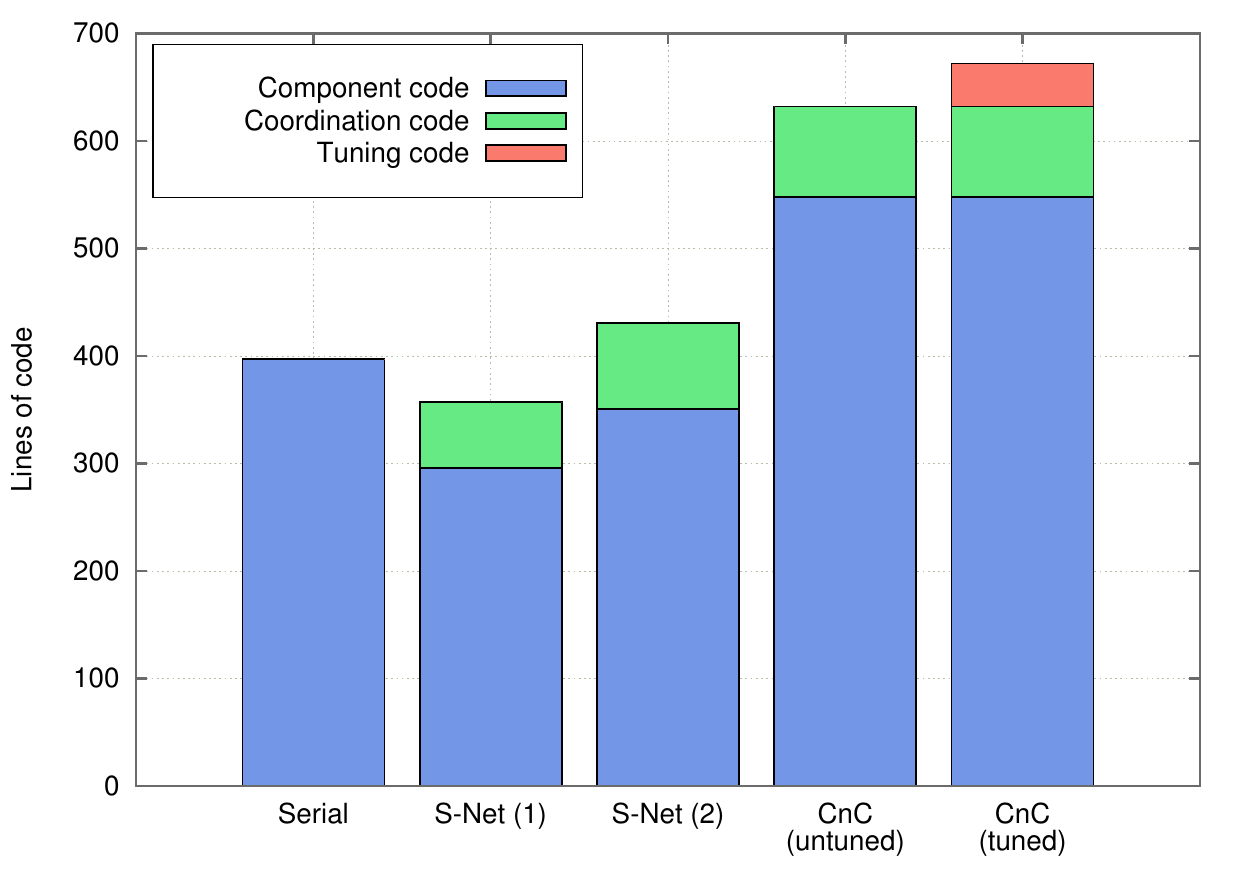}
       {!t}
       {width=3.3in}
       {Number of lines of code for each implementation.}
       {pl:lines}

\subsubsection{Discussion}

In terms of expressiveness, all CnC and \snet{} implementations are
well-structured and there is a reasonably low amount of extra code required to
pay for coordination. In most cases, the implementation consists of two parts:
the high-level structure of the program represented as a graph and a set of
functions defining box/step behavior that use an API for interacting with
\snet{}/CnC respectively. In addition, tuning code may be present in the
CnC program.

We demonstrate the programmability of each approach by showing the number of
lines of code for each program in Fig.~\ref{pl:lines}.  It gives a rough idea
of the amount of coordination and tuning overhead.  Overall,
the amount of coordination code is relatively small in either case.  CnC
components are implemented in C++ and representing each component as a separate
class requires some additional overhead.

To summarize, CnC and \snet{} are both systems for coordination of components.
Components in CnC are linked by data and control relations.  The execution
strategy and computation order are determined mainly dynamically.  \snet{} uses
a message-driven strategy.  Here components are linked by typed data relations
only.  Additionally \snet{} offers a hierarchical structuring mechanism and
a compile time analysis which supports this structure; most of that analysis
can be done statically.

\section{Performance Measurements}

Cholesky decomposition is a significant example of a computational linear
algebra problem. It is affected by both locality (tile size) and parallelism
(the number of cores used). The amount of work and available parallelism varies
during the run time,  which should reveal the differences in both systems'
abilities to manage the resources of a concurrent platform.

The CnC model permits many run-time system designs, including those for
distributed memory systems using MPI as well as shared memory versions. We use
Intel CnC 0.9, which uses Intel Thread Building Blocks (TBB) as a threading
layer.  Step components are implemented in C++ and compiled into libraries
using GNU GCC 4.6.3.

Boxes for \snet{} were implemented in \langc{} and were compiled with GNU GCC
4.6.3.  All measurements were done with the \front{} runtime system
\cite{Gijsbers13,GijsGrelIJPP14}. 
This is a novel runtime system, which combines a very low
overhead of \snet{} box/network instantiation with efficient transportation of
records throughout the network and box replication. 
Experiments show that this runtime system scales well
to very large \snet{} networks with millions of concurrent records.

All measurements are obtained on a machine with four twelve-core AMD
Opteron\textsuperscript{\texttrademark} 6174 Processors and 256 GB RAM (see
Table~\ref{tab:hardware} for technical details). 
When measuring speedup for increasing core counts, we first employ 
neighboring cores in the same processor before going beyond processor
boundaries.

\begin{table}[!t]
\caption{Evaluation platform for experiments}
\begin{center}
\begin{tabular}{c|c}
\hline
Vendor & AMD\\
Processor\ Model & Opteron 6174\\
Processor\ Name & Magny-Cours\\
\hline
Clock (GHz) & 2.2\\
\# Sockets & 4\\
Cores(Threads)/Socket & 12(12)\\
L1 Data Cache & 64 KB/core\\
L2 (Data and Instruction) Cache & 512 KB/core\\
Shared L3 Cache & 12 MB\\
DRAM Capacity & 256 GB\\
\hline
\end{tabular}
\end{center}
\label{tab:hardware}
\end{table}

The serial implementation was used as a reference to measure the speedup.  This
version implements the tiled Cholesky decomposition algorithm for a single core
without use of optimized kernels.  It is similar to the implementation in CnC,
but without any coordination overhead.  The program consists of three essential
functions, each representing a single step of the algorithm shown in
Fig.~\ref{alg:cholesky}.  As the algorithm states, these functions are called
in a loop with different parameters.  During each iteration a global array of
tiles is accessed to read or to write tiles.  This is similar to how the data
collection is accessed in CnC.

\pdffig{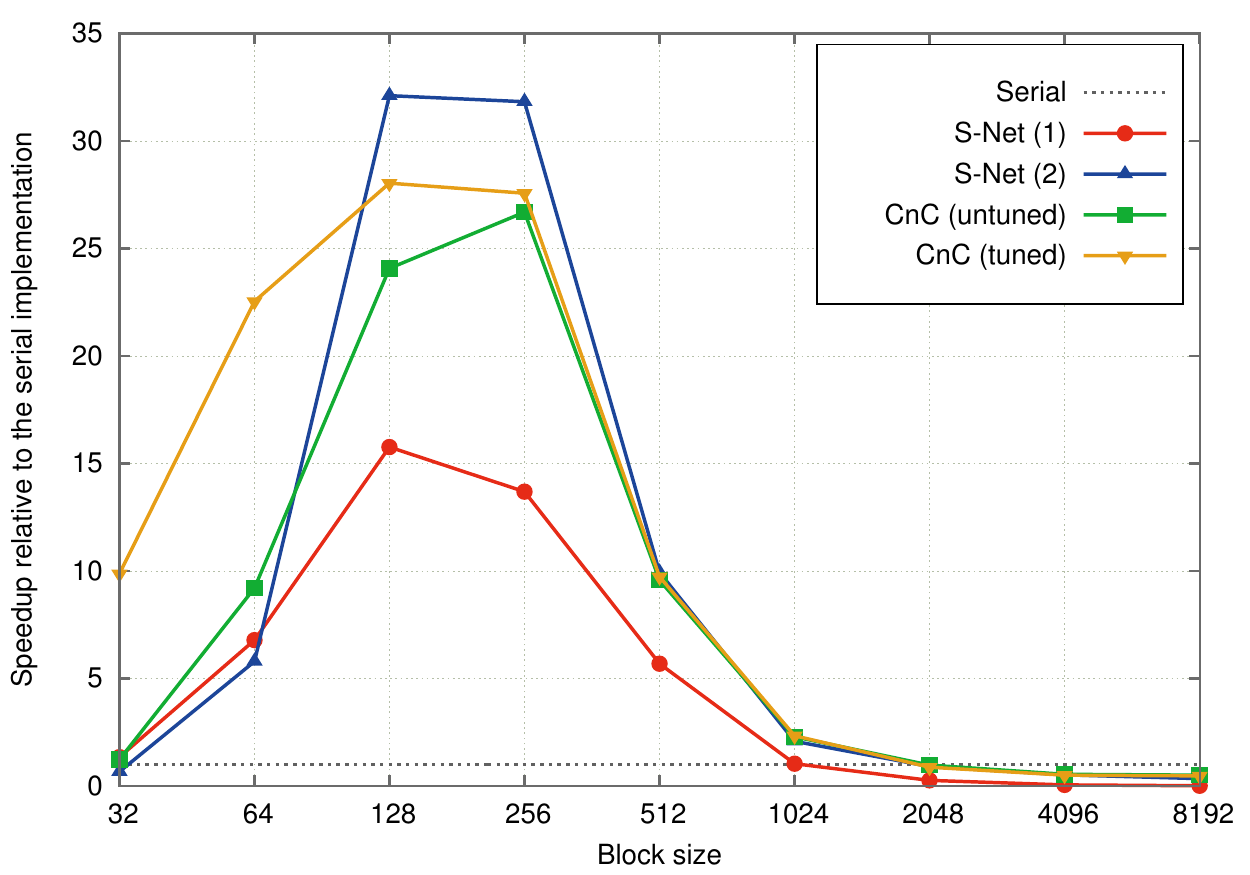}
       {!t}
       {width=3in}
       {The speedup relative to the serial implementation of Cholesky
        decomposition CnC and \snet{} applications on 48-core machine for
        an input matrix of size $N=8192$.}
       {pl:tiles_speedup}

\pdffig{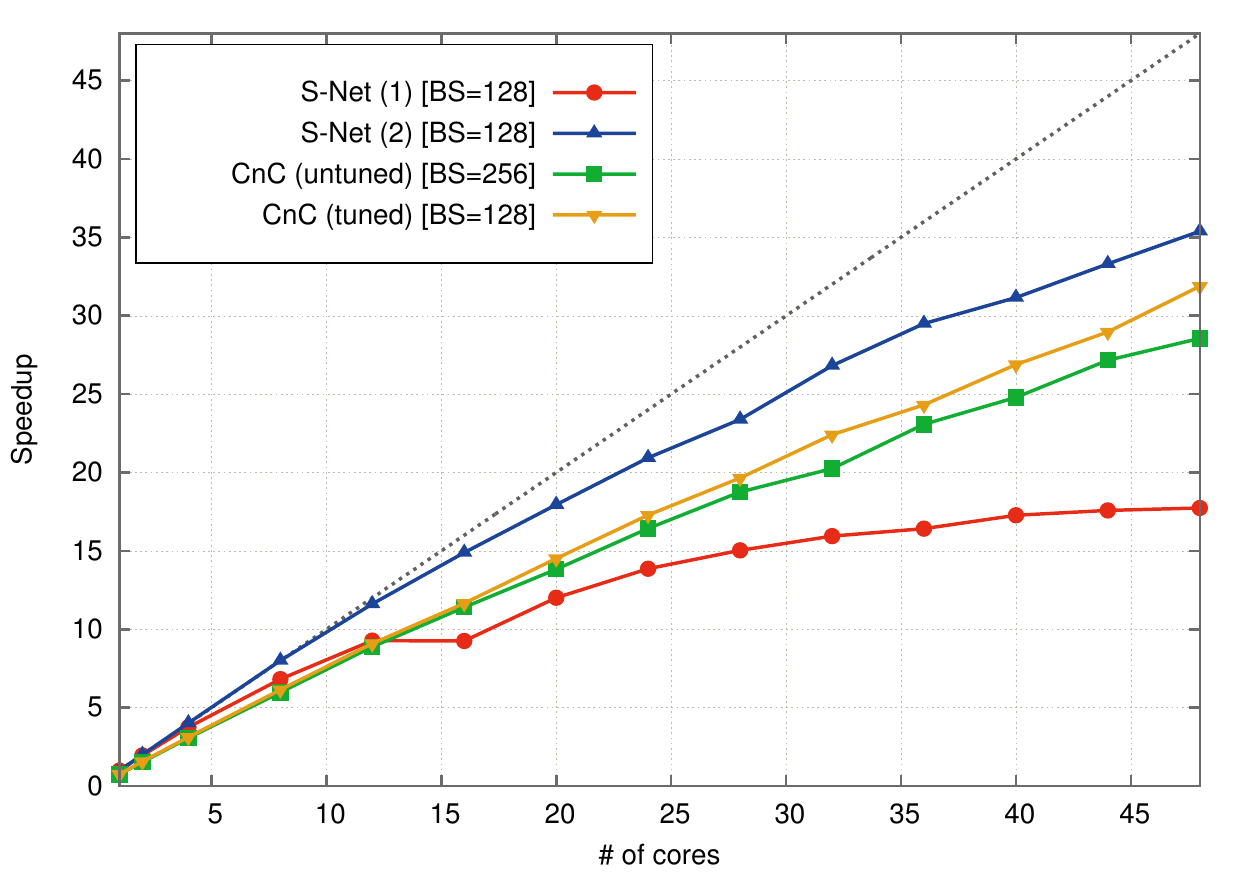}
       {!t}
       {width=3in}
       {The speedup of Cholesky decomposition CnC and \snet{} applications for
        different number of cores (for matrix of size $8192 \times 8192$
        and optimal block size).}
       {pl:cores_speedup}

We provide evaluation results for the two CnC and the two \snet{} implementations
of tiled Cholesky factorization as described in Section~\ref{sec:case-study}.
Fig.~\ref{pl:tiles_speedup} shows the speedup relative to the serial
implementation executed with optimal block size $b=128$ as a parameter.  
We systematically vary block sizes for both CnC and \snet{} implementations.
In this figure the overhead can be observed for large blocks, where the speedup
of concurrent versions drops significantly.  Four peaks in the figure
demonstrate optimal values for block size.  The maximum speedup is 32 that was
achieved by \snet{} data-driven implementation.

The use of dependency functions in the tuned CnC implementation brings
significant improvement compared to the untuned version of the program,
especially when there is a high amount of concurrency (for Cholesky
decomposition it is the case where the size of subproblems is small).  In the
best case it brings almost a factor of 8 improvement for the current
application.  For the best performance range of tile size improvement caused by
dependency functions is about 15\%.  On the other hand, an overhead introduced
by the dependency functions for cases with a small amount of concurrency caused
a 7\% loss in performance.

Fig.~\ref{pl:cores_speedup} shows speedups of \snet{} and CnC for increasing
numbers of cores.
As pointed out before, we first use all cores of one processor before proceeding
to the next processor when increasing the effective core count
to exploit the memory hierarchy.
All applications were executed with optimal block sizes.
Within the field of tested implementations the data-driven \snet{} version
of tiled Cholesky factorization achieves the best performance and scales up
to the maximum of 48~cores. 
Both CnC implementations are clearly behind, although they likewise demonstrate
excellent scalability. 
Making use of the tuning facilities of CnC results in a rather marginal
performance advantage in this experiment.
As expected, the \snet{} implementation involving barrier synchronization
suffers from lesser scalability and increasing overhead as the number of
cores used grows. 
We can, in particular, observe this effect when moving from 12~cores to
16~cores, i.e.~beyond a single processor.

A thorough performance evaluation of the tiled Cholesky decomposition in CnC
can be found in \cite{ChanKnobVuduIPDPS10}.  The results of the measurements
illustrate that \snet{} on this example has a performance similar to CnC and
that coordination programming model is an effective instrument for implementing
applications for multi-core platforms.

The second \snet{} implementation clearly benefits from the dataflow model
which \snet{} provides.  The execution of a component is enabled by the arrival
of input data.  This paradigm allows for the specification of highly-concurrent
applications.  In contrast, execution of CnC component is prescribed by tag
without awareness of data availability.  This may introduce stalls during
execution.   A significant speedup in CnC was achieved by introducing
dependency functions, which map tag indices to element indices in data
collection.

\section{Conclusions}
\label{sec:conclusions}

We presented a performance case study on a popular linear algebra problem using coordination
programming as a method of code development.  We compared two design styles of coordination
programming and their runtime performance on a large multicore server:
our coordination language \snet{} vs Intel's coordination library/specification
tool CnC (Concurrent Collections).

We observe that a static network topology and data relations facilitates
\snet{} compilation and run-time scheduling and communication.  \snet{} does
not use control flow, allowing components to be triggered merely by the
availability of their input data.  Despite the lack of tag collections that
determine the sequencing of processing steps, pure dataflow works quite well,
outperforming CnC in the best performance range of problem sizes.  \snet{}
supports a clean separation of concerns between coordination and computation:
only individual objects required by a computational component are delivered to
it by a coordinator.  By contrast, in CnC components must be aware of the whole
data collections they wish to access.

Tuning is a feature of CnC that is clearly separated from application design.
By introducing ``depends'' functions to the application we demonstrated the
improvement this can bring.  For the current application it delivers an
8 fold improvement in the best case.

Both of CnC and \snet{} are designed to maximize programmability and usability
of various many-core platforms.  We compared the two coordination models with
a serial implementation.   We managed to achieve optimal utilization of the
resources without platform-specific tuning and optimization.  The data-driven
implementation is \snet{} is based on precisely the same sequence of
algorithmic steps as the CnC one (though implementation with a barrier
synchronization is different).  In order to increase the performance for CnC,
one should consider other features of the tuning mechanism (i.e.~priorities)
that may improve scheduling and memory management at the run-time.

\section*{Acknowledgment}
This work has made use of the University of Hertfordshire Science and
Technology Research Institute high-performance computing facility.
The authors further wish to thank the anonymous reviewers for their
valuable comments.